\documentclass[pra,aps,showpacs,groupedaddress,superscriptaddress,twocolumn,longbibliography]{revtex4-1}

\usepackage[utf8x]{inputenc}
\usepackage{color}
\usepackage{bbm}

\usepackage{nicefrac}
\usepackage{amsfonts,amsmath,amssymb,stmaryrd}
\usepackage{braket}
\usepackage[english]{babel}

\usepackage{tabularx}
\usepackage{multirow} 
\usepackage{array}  
\usepackage{hhline}

\usepackage{graphicx}
\usepackage{grffile}
\usepackage{subfigure}  
\usepackage{bm}	
\usepackage{bbm}	
\usepackage{hyperref}
\usepackage[pdftex]{epsfig}
\usepackage{mathrsfs}
\usepackage{verbatim}
\usepackage{centernot}

\usepackage{cancel,ifthen}

\usepackage{todonotes}
\presetkeys{todonotes}{size=\tiny}{}

\InputIfFileExists{gitinfo-latex.inc}{}{}
\usepackage[acronym,nomain]{glossaries}

\usepackage{units}
\usepackage{upgreek}

\renewcommand{\vec}[1]{\mathbf{#1}}

\def\qc1{q_c^{(1)}}

\def\v0{\vec{0}}

\def\gib{g_\text{IB}}

\def\mI{m_{\rm I}}  
\def\mB{m_{\rm B}}  

\renewcommand{\l}{\left(}
\renewcommand{\r}{\right)}

\renewcommand{\vec}[1]{\bm{#1}}

\newcommand{\cmnt}[2][NoInPuT]{\ifthenelse{\equal{#1}{NoInPuT}}{}{{\color{red}\sout{#1}}} {\color{blue} #2}}

\begin{document}
	
	\title{Rotational cooling of molecules in a Bose-Einstein-Condensate}
	
	\author{Martin Will}
	\author{Tobias Lausch}
	\author{Michael Fleischhauer}
	\affiliation{Department of Physics and Research Center OPTIMAS, University of Kaiserslautern, 67663 Kaiserslautern, Germany}

	
	\date{\today}
	

	\begin{abstract}
		We discuss the rotational cooling of diatomic molecules in a Bose-Einstein condensate (BEC) of ultra-cold atoms by emission of phonons with orbital angular momentum. Despite the superfluidity of the BEC there is no frictionless rotation for typical molecules 
		since the dominant cooling occurs via emission of particle-like phonons. 
		Only for macro-dimers, whose size becomes comparable or larger than the condensate healing length, a Landau-like, critical angular momentum exists below which phonon emission is suppressed. 
		We find that the 
		rotational relaxation of typical molecules is in general faster than the cooling of the linear motion of impurities in a BEC. This also leads to a finite lifetime of angulons, quasi-particles of rotating molecules coupled to phonons with orbital angular-momentum. We analyze the dynamics of rotational cooling for homo-nuclear diatomic molecules based on a quantum Boltzmann equation including single- and two-phonon scattering and discuss the effect of thermal phonons.
	\end{abstract}
	
	\maketitle
	

	\section{Introduction}
	
\begin{figure}[htb]
		\begin{center}
			\includegraphics[width=0.48\textwidth]{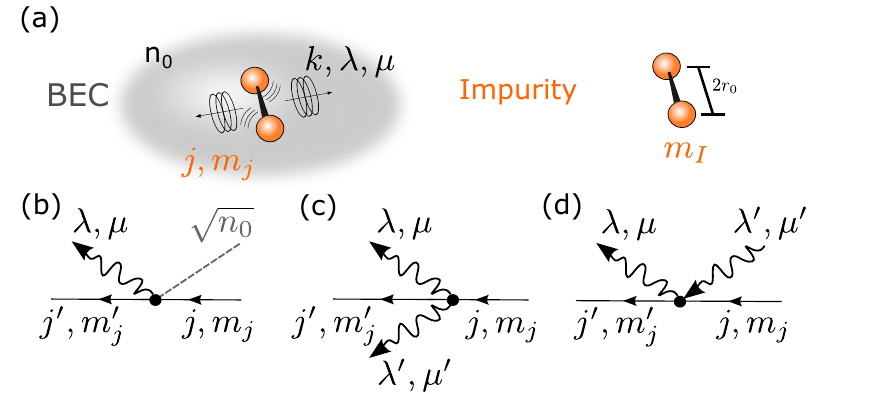}
		\end{center}
		\caption{(Color online) a) A rotating dimer with mass $m_I$, diameter $2r_0$ and rotational quantum numbers $j, m_j$ immersed into a BEC of density $n_0$ undergoes rotational relaxation by emission of phonons with orbital angular momentum and quantum numbers $k, \lambda,\mu$.  b) Spontaneous creation of phonons by
			interaction with condensate atoms. The inverse process requires the availability of thermal excitations. c) Spontaneous or thermally induced two-phonon creation. ($\prec$) d)  Scattering and exchange of angular momentum between excitation and impurity dimer ($\times$). 
		}
		\label{fig:Setup}
	\end{figure}

	The physics of a quantum impurity in collective many-body environments is an important subject of condensed matter 
	physics.  It dates back to the classic problem of a polaron put forward by Landau and Pekar
	\cite{Landau1948}  and Fr\"ohlich and Holstein \cite{Froehlich1954,Holstein-1959-a,Holstein-1959-b,Devreese-RPP-2009} to explain charge transport in solids resulting from 
	the dressing of a moving electron with phonon-like excitations of the surrounding material. 
	In many systems internal degrees of freedom of the impurity can be disregarded as their characteristic energy scale
	is well separated from that of the environment and the impurity can be treated as point-like object.
	The Fermi and Bose polarons recently realized in ultra-cold quantum gases \cite{Prokofev-PRB-2008,Nascimbene-PRL-2009,Schirotzek-PRL-2009,
		Koschorrek-Nature-2012,Rath2013,Scelle-PRL-2013,Levinsen2015,Ardila2015,Hu-PRL-2016,Jorgensen-PRL-2016,Shchadilova2016,Sidler-NatPhys-2017} are important examples providing 
	a many-body model system where impurity problems can be analyzed very precisely. 
	Also
	the dynamics of its formation can be studied, which is an equally important problem since collective properties such as the 
	superfluidity of a BEC can strongly influence the equilibration dynamics \cite{SchmidtPRL2018,Lausch2018a,Lausch2018b}.
	Recently the concept of a polaron was extended to impurities with a more complex structure
	such as a molecule. It was shown that the coupling of 
	rotation to collective excitations of a surrounding BEC can give rise to a new type of quasi-particles termed angulons \cite{Schmidt2015,Schmidt2016,Lemeshko2017,Lemeshko-PRL-2017,Bighin-PRL-2018}.
	In the present paper we discuss the cooling dynamics of the rotational degrees of freedom of
	a single, diatomic molecule immersed in a three-dimensional (3D) Bose-Einstein condensate, see Fig.~\ref{fig:Setup}, which
	is relevant both for the formation and the stability of angulons. To this end we use a microscopic quantum Boltzmann approach \cite{Greiner1998} based on
	a Bogoliubov theory of impurity-condensate interaction.

	Emission and scattering of Bogoliubov phonons with orbital angular momentum off the molecule lead to a deceleration of the rotational motion and eventually to equilibration with the condensate. 
	For typical sizes of molecules and weakly interacting condensates  there is no analogue of a Landau critical velocity, i.e. there is
	in general no critical value of angular momentum below which phonon emission and scattering is suppressed. This is because the spatial structure of the molecule
	can only be resolved by high-energy phonons, which have a particle-like character. Thus different from the case of polarons, i.e. point-like impurities dressed with Bogoliubov phonons, there are in general no stable states of angulons. The rotational relaxation rates are however smaller than the typical binding energies of angulons. 
	
	
	The situation is different if one considers macro-dimers, such as Rydberg molecules \cite{Bendkowsky2009, Tallant2012, Anderson2014, Greene2000}, where an
	atom is trapped in a high-lying Rydberg state of another atom. In this case molecular size and healing length can become comparable and the interaction with low-energy phonons becomes the most important one. The same holds true for impurities trapped in shallow, rotationally symmetric potentials.
	In this limit the superfluidity of the condensate changes the relaxation dynamics and we recover a Landau critical behaviour. Below a certain angular momentum of the macro-dimer the emission of phonons is
	effectively suppressed and the rotational relaxation stops in a pre-thermalized state.
	
	The paper is organized as follows: In Sec.\ref{sec:model} we will introduce the model of a rigid rotor coupled to Bogoliubov phonons of an atomic BEC.
	The quantum Boltzmann equation used to describe the relaxation dynamics is reviewed in Sec.\ref{sec:QBE} and the different contributions to the relaxation rates
	resulting from spontaneous and thermal single- and two-phonon processes are derived. The relaxation dynamics of macro-molecules will be discussed in
	Sec.\ref{sec:Landau} and that of typical molecules in Sec.\ref{sec:small}.

	\section{Model}
	\label{sec:model}
	
	We here discuss the case of a diatomic molecule, which we describe as a rigid rotor of two point masses $m_I$ with distance $2 r_0$, see Fig.\ref{fig:Setup}, immersed in
	a three-dimensional (3D) weakly interacting Bose Einstein condensate of atoms, which we describe in Bogoliubov approximation.  We assume that the center of mass (COM) of the molecule is at rest in the lab frame of the BEC and we disregard the COM kinetic energy of the molecule. 	
	The total Hamiltonian
	\begin{equation}
	H=H_0^\textrm{m}+H_0^\textrm{ph}+H_\textrm{int}
	\end{equation}
	consists of the free Hamiltonians of the diatomic molecule $H_0^\textrm{m}$, the interaction $H_\textrm{int}$ and that of the Bogoliubov phonons  $H_0^\textrm{ph}$\cite{Pitaevskii2016}:
	\begin{equation}
	H_0^\textrm{m}=\frac{\hat L^2}{4m_I r_0^2},\qquad H_0^\textrm{ph} = \sum_{k,\lambda,\mu} \omega_k \, \hat b_{k\lambda\mu}^\dagger \hat b_{k\lambda\mu}.
	\end{equation}
	Here $\hat L$ is the angular momentum operator of the rotating diatomic molecule and $2 r_0$ the molecule diameter.
	$\omega_k=c k\sqrt{1+k^2\xi^2/2}$ is the Bogoliubov dispersion relation of phonons with momentum $k$, with $\xi = 1/\sqrt{2 m_B g n_0}$ being the condensate healing length. $m_B$ is the mass of the BEC atoms, $g$ is the strength of
	atom-atom interactions in the condensate in $s$-wave approximation. $c=\sqrt{ g n_0/m_B}$ is the speed of sound of the phonons. 
	The homogeneous condensate of density $n_0$ is assumed to be in an initial equilibrium state at temperature $T\ll T_c$, $T_c$ being the critical temperature of condensation, which for a non-interacting homogeneous condensate of density $n_0$ reads $T_c=2\pi n_0^{2/3}/(m_B\zeta(3/2)^{2/3})$. If a rotating molecule is placed in the BEC we expect that its angular momentum thermalizes to an equilibrium distribution of quantum numbers $j$ with characteristic value,
	\begin{equation}
	j_T(j_T+1) = \frac{8\pi}{\zeta(3/2)^{2/3}}\, \frac{T}{T_c}\,\frac{m_I}{m_B}\Bigl(r_0 n_0^{1/3}\Bigr)^2,
	\end{equation}
	For a typical molecule with size small compared to the average distance between atoms in the BEC, i.e. $r_0\ll n_0^{-1/3}$,
	we expect a cooling to the lowest angular momentum $j\to 0$. 

	The interaction of the homo-nuclear diatomic molecule with the BEC, $H_\textrm{int} = H_\textrm{I}(\mathbf{r})+  H_\textrm{I}(-\mathbf{r})$, is described as $s$-wave scattering interaction of the two atoms with the condensate. We assume that higher-order partial waves are not relevant for the scattering process process with small rotational quantum numbers. They result in modifications of the dispersion relation that has been discussed e.g. for He-dimers in \cite{Lemeshko2017}. The interaction $H_\textrm{I}(\mathbf{r})$ of a point-like impurity at position $\mathbf{r}$ with the BEC reads in terms of plane-wave Bogoliubov modes
	\begin{eqnarray}
	H_\textrm{I}(\mathbf{r}) &=& \int \!\! d^3 k   \frac{\gib n_0^{1/2}}{(2 \pi)^{3/2}}  W_k e^{-i \vec{k} \cdot {\vec{r}}} \l \hat{ 
		b}^\dagger_{\vec{k}} + \hat {b}_{-\vec{k}}  \r \nonumber\\ 
	&&+\frac{\gib}{2 (2 \pi)^3}  \int \!\! d^3k \int \!\! d^3k'   \biggl[ W^{\times}_{k,k'} {\hat b}^\dagger_{\vec{k}} {\hat b}_{\vec{k}'}+\label{eq:HBogoPolaron} \\
	&&\qquad+ \frac{1}{2} W^{\prec}_{k,k'}  \l {\hat b}^\dagger_{\vec{k}} {\hat b}^\dagger_{-\vec{k}'} + {\hat b}_{-\vec{k}} {\hat b}_{\vec{k}'} \r \biggr] e^{-i \l \vec{k} - \vec{k}' \r \cdot {\vec{r}}}.\nonumber
	\end{eqnarray}
	where $W_k =[k^2 \xi^2 / (2 + k^2 \xi^2)]^{1/4}=\sqrt{\epsilon_k/\omega_k}$, with $\epsilon_k = k^2/(2 m_B)$ being the kinetic energy of the
	condensate atoms, and we used the abbreviations $W^{\times}_{k,k'}= W_k W_{k'}+W^{-1}_k W^{-1}_{k'}$ and 
	$W^{\prec}_{k,k'}= W_k W_{k'}-W^{-1}_k W^{-1}_{k'}$. 
	
	Making use of the decomposition of plane waves into spherical ones
	\begin{eqnarray*}
		e^{i\vec{k}\cdot {\vec{r}} } &=& 4\pi \sum_{\lambda\mu} i^\lambda\, j_\lambda\left( k{r}\right)\, Y_{\lambda\mu}({\theta},{\phi})\, Y_{\lambda\mu}^{*}(\theta_k,\phi_k)
	\end{eqnarray*}
	where $j_\lambda(kr)$ is the spherical Bessel function, and the orthogonality relations of spherical harmonics, we can rewrite eq.(\ref{eq:HBogoPolaron}) in terms of angular momentum modes
	\begin{eqnarray*}
		\hat b_{k\lambda\mu} &=& k \int\!\! d\phi_k\!\int \!\! d\theta_k \sin\theta_k i^\lambda \,Y^{*}_{\lambda\mu}(\theta_k,\phi_k)\, \, \hat b_\textbf{k}	,\\
		\hat b_\textbf{k} &=& \frac{1}{k}\sum_{\lambda\mu} i^{-\lambda} \, Y_{\lambda\mu}(\theta_k,\phi_k)\,\, \hat b_{k\lambda\mu}.
	\end{eqnarray*}
	$\lambda = 0,1,\dots$ and $\mu=-\lambda,-(\lambda-1),\dots,\lambda-1,\lambda$ are the quantum numbers of the orbital angular momentum of the phonons in the rest frame of the center-of-mass of the molecule. The spherical-mode operators fulfill bosonic commutation relations $[\hat b_{k\lambda\mu} ,\hat b_{k^\prime\lambda^\prime\mu^\prime}^\dagger] = \delta(k-k^\prime) \delta_{\lambda,\lambda^\prime}\delta_{\mu,\mu^\prime}$. With this we find
	\begin{eqnarray}
	&&H_\textrm{int} =
	\sum_{k \lambda \mu} U_\lambda(k)\; 
	\left[
	Y_{\lambda\mu}(\theta,\phi) \; \hat{b}_{k\lambda\mu} + Y_{\lambda\mu}^*(\theta,\phi) \; \hat{b}^\dagger_{k\lambda\mu}
	\right] +\nonumber\\
	&& \enspace+ 
	\sum_{\substack{k\lambda\mu \\ k'\lambda'\mu'}}
	U^\times_{\lambda\lambda'}(k,k') Y_{\lambda'\mu'}^*(\theta,\phi)Y_{\lambda\mu}(\theta,\phi)\,\, \hat{b}^\dagger_{k'\lambda'\mu'} \hat{b}_{k\lambda\mu}+
	\label{eq:Hint}\\
	&& \enspace+
	\sum_{\substack{k\lambda\mu \\ k'\lambda'\mu'}} 
	\frac{1}{2} U^\prec_{\lambda\lambda'}(k,k')\; 
	Y_{\lambda\mu}(\theta,\phi)Y_{\lambda'\mu'}(\theta,\phi) \,\, \hat{b}_{k'\lambda'\mu'} \hat{b}_{k\lambda\mu}+ h.a. \nonumber
	\end{eqnarray}
	where we made use of the fact that the distance of both atoms to the origin is the same and fixed to $r=r_0$.
	The coupling constants for the single-phonon terms read
	\begin{eqnarray}
	U_\lambda\ &=&
	\begin{cases}
	\, g_\text{IB} \, \sqrt{\frac{8n_0}{\pi}}\,k\,W_k \; j_{\lambda}(k r_0) & \quad \lambda \enspace\text{even}\\
	0 & \quad \lambda \enspace \text{odd}
	\end{cases}  
	\label{DefU}
	\end{eqnarray}
	and for the two-phonon terms
	\begin{eqnarray}
	U^\mu_{\lambda\lambda'} &=&
	\begin{cases} 
	\frac{2 g_\text{IB}}{\pi} k k' \;
	j_{\lambda}(k r_0) \; j_{\lambda'}(k' r_0)\,
	W^{\mu}_{k,k'} &  \lambda+\lambda' \text{even\qquad}\\
	0 &   \lambda+\lambda' \text{odd}
	\end{cases} 
	\label{DefU2}
	\end{eqnarray}
	The vanishing of the coupling constants for odd values of $\lambda$ or $\lambda+\lambda'$ 
	is due to the inversion symmetry of the molecules. For the hetero-nuclear case also odd terms would be
	nonzero. As a consequence the symmetric molecule can only emit and absorb single phonons with even orbital angular momentum
	or phonon-pairs which have an even total angular momentum. Rotational cooling will thus occur in a cascade with angular momentum 
	steps of two.
	
	\section{Quantum Boltzmann equation}
	\label{sec:QBE}
	
	We now want to study the dynamics of a molecular impurity with finite initial angular momentum interacting with the BEC, described by the Hamiltonian (\ref{eq:Hint}). 
	The starting point is  a master equation for the impurity-density matrix,  $\rho_{jj'}^{mm'}$ between angular momentum states which can 
	be derived by integrating out the phonon degrees of freedom and employing a Born-Markov approximation. The Born approximation neglects higher-order scattering contributions and is valid for weak impurity-condensate interactions $g_\textrm{IB}$. On a short time  scale
	off-diagonal matrix elements dephase and it is sufficient to consider probabilities $p_{jm}=\rho_{jj}^{mm}$ only, for which we obtain a
	linear Boltzmann equation, with transition rates obeying Fermi's golden rule $\Gamma_{m\to n}= 2\pi \delta(E_m-E_n) \vert \langle m\vert H_\textrm{int} \vert n\rangle \vert^2$  \citep{Greiner1998}.
	\begin{equation}
	\frac{dp_{jm}}{dt\,\,}
	= \sum_{j'm'} \Bigl( p_{j'm'} \Gamma_{j'm' \to jm} \,-\,p_{jm} \Gamma_{jm \to j'm'}\Bigr)\label{eq:Boltzmann}
	\end{equation}
	In the following we will derive the transition rates resulting from single- and two-phonon processes.
	
	\subsection{Single-phonon transition rates}
	
	In order to determine the transition rates $\Gamma_{jm \to j'm'}$ from (\ref{eq:Hint}), we make use of the matrix elements of spherical harmonics
	\begin{eqnarray*}
		&&\langle j,m\vert Y_{\lambda\mu}(\theta,\phi)\vert j',m'\rangle =
		\sqrt{\frac{1}{4\pi}} \, G^{jm}_{j'm',\lambda\mu},\nonumber\\
		&&G^{jm}_{j'm',\lambda\mu}= 
		\sqrt{\frac{(2j'+1)(2\lambda+1)}{(2j+1)}} \; C_{j'm',\lambda\mu}^{jm} \;	C_{j'0,\lambda0}^{j0},
	\end{eqnarray*}
	where $C_{jm,j'm'}^{\lambda\mu}$ are Clebsch-Gordan coefficients, which reflect angular momentum conservation.
	
	As discussed in detail in the Appendix the spontaneous (sp) and thermal (T) contributions resulting from the single-phonon term in the interaction hamiltonian read:
	\begin{eqnarray}
	\Gamma^\textrm{1ph,sp}_{jm \to j'm'} &=&
	\sum_{\lambda\mu}  \gamma_{\lambda}^{jj'} G^{jm^2}_{j'm',\lambda \mu} \Theta_{j,j'} \label{eq:single-phonon_m_resolved_sp},	 \\
	\Gamma^\textrm{1ph,T}_{jm \to j'm'} &=& 
	\sum_{\lambda\mu}  \gamma_{\lambda}^{jj'} G^{jm^2}_{j'm',\lambda \mu}  \overline{n}_{jj'} \label{eq:single-phonon_m_resolved_th}.
	\end{eqnarray}
	$\Theta_{j,j'} = \Theta(j-j')$ is the heaviside step function and $\overline{n}_{jj'}=(\exp\{\vert E_{jj'}\vert /k_B T\}-1)^{-1}$ is the thermal phonon number corresponding to the transition energy $E_{jj'} = E_j-E_{j'}$ between rotational states with $E_j= j(j+1)/m_I r_0^2$. The effective transition rates for angular-momentum transfer $\lambda$ are given by 
	\begin{align}
	\gamma_\lambda^{jj'}=
	\begin{cases}
	\frac{\displaystyle{4 g_\text{IB}^2 n_0}}{\displaystyle{\sqrt{2}c \xi^2\pi}}
	\frac{\displaystyle{\bigl(k_{jj'}\xi\bigr)^3}}{\displaystyle{\sqrt{1+2 E_{jj'}^2 \xi^2/c^2}}}\,  j_\lambda^2 
	\Bigl( r_0 k_{jj'}\Bigr)
	\quad &\lambda\enspace \text{even }\\
	0 &\text{else}
	\end{cases}
	\label{eq:effective one phonon rates}
	\end{align}
	where $k_{jj'}$ is the phonon momentum corresponding to $E_{jj'}$.  
	
	The discussion can be substantially simplified if we consider only the total probabilities for angular momentum $j$,
	$p_j = \sum_{m=-j}^j p_{jm}$. Making use of the properties of Clebsch-Gordan coefficients we find that the total rates $\sum_{m'} \Gamma_{jm \to j'm'} =\Gamma_{j\to j'}$ are independent of $m$ as expected from the rotational symmetry of the problem. Thus eq.(\ref{eq:Boltzmann}) simplifies to
	\begin{equation}
	\frac{dp_{j}}{dt\,\,}
	= \sum_{j'} \Bigl( p_{j'} \Gamma_{j'\to j} \,-\,p_{j} \Gamma_{j \to j'}\Bigr)
	\label{eq:Boltzmann-simple}
	\end{equation}
	with the total rates
	\begin{eqnarray}
	\Gamma_{j\to j'}^\textrm{1ph,sp} &=&
	(2j'+1) \sum_\lambda \gamma_\lambda^{jj'} \left(C_{j0,j'0}^{\lambda0}\right)^2\,  \Theta_{j,j'} ,\nonumber\\
	\Gamma_{j\to j'}^\textrm{1ph,T} &=& 
	(2j'+1)\sum_\lambda \gamma_\lambda^{jj'} \left(C_{j0,j'0}^{\lambda0}\right)^2\,  \overline{n}_{jj'}. 
	\label{eq:single-phonon-rates}
	\end{eqnarray}
	Since for the Clebsch-Gordan coefficients holds $C_{j0,j'0}^{\lambda0}=0$, if $j'+j+\lambda$ is odd and $\gamma_\lambda^{jj'}=0$ for
	odd $\lambda$, one recognizes that states with even (odd) initial angular momentum $j$ can only decay into states with even (odd) final angular momentum $j'$.
	
\begin{figure}[htb]
		\begin{center}
			\includegraphics[width=0.5\textwidth]{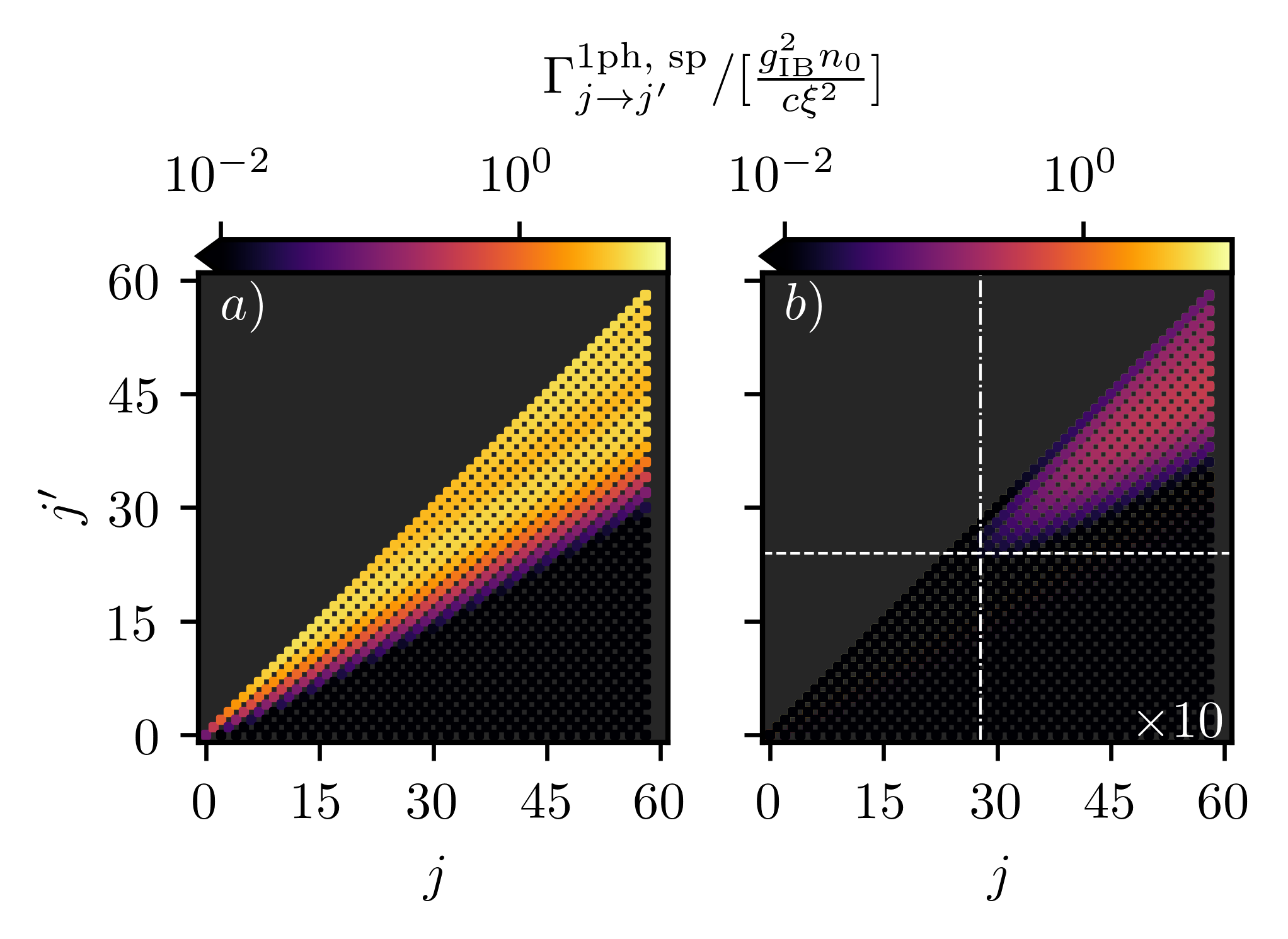}
		\end{center}
		\caption{(Color online) Spontaneous single-phonon decay rates $\Gamma_{j\to j'}^\textrm{1ph,sp}$ as function of angular-momenta $j$ and $j'$ for normal molecule sizes, $r_0=0.1\xi$ shown in (a), and a macro-dimer with $r_0 = 10 \xi$, shown in (b) amplified by a factor of 10. In both cases the mass ratio is $m_I = 2 m_B$. Dark gray background means zero scattering amplitude. In (b) the decay rates are suppressed for a initial angular momenta $j < j_c$(vertical line) and there is no scattering below $j' < j_c^{(1)}$ (horizontal line), eq.\eqref{eq:jc} and \eqref{eq:jc1}. 
		}
		\label{fig:single-phonon}
	\end{figure}

	In order to get an impression of the dependence of the single-phonon decay rates on the angular momentum quantum numbers, we have plotted in Fig.~\ref{fig:single-phonon} the spontaneous scattering rates $\Gamma_{j\to j'}^\textrm{1ph,sp}$ as functions of $j$ and $j'$ for
	two different values of $r_0/\xi$. While for typical sizes of molecules, for which $r_0/\xi\ll 1$, shown in  Fig.~\ref{fig:single-phonon}(a), there is
	a smooth dependence on $j$ and $j'$, one finds for macro-dimers, for which $r_0/\xi$ is of the order of or larger than unity, shown in  
	Fig.~\ref{fig:single-phonon}(b), that the decay rates are strongly suppressed for $j$ below a critical value $j_c$. 
	Also the final angular momentum that can be reached in a single-phonon process is limited by a second critical value $j_c^{(1)}$.
	This will be discussed in more detail in sec.\ref{sec:Landau}. 
	
	\subsection{Two-phonon transition rates}
	
	For the calculation of the two-phonon transition rates we need the matrix elements of the product of spherical harmonics
	\begin{eqnarray}
	&&\bra{j'm'} Y_{\lambda'\mu'}(\theta,\phi)Y_{\lambda\mu}(\theta,\phi )\ket{jm}= \notag \\
	&&\qquad \qquad= \frac{1}{4 \pi}  \sum_{LM} G^{LM}_{\lambda'\mu' , \lambda \mu} G^{j'm'}_{jm ,LM}.
	\end{eqnarray} 
	As discussed in the Appendix and shown in Fig.\ref{fig:Setup} we find for the total transition rate corresponding to the scattering of a phonon off the molecule ($\times$) and the simultaneous excitation of two phonons ($\prec$), described by the
	two-phonon interaction terms in eq.(\ref{eq:Hint})%
	\begin{eqnarray}
	\Gamma^\times_{j\to j'} &=& 
	(2j'+1)\!
	\sum_{
		\substack{L \text{ even} \\
			\lambda \, \lambda' 
	}}\!\!
	\frac{(2\lambda+1)(2\lambda'+1)}{2L+1}
	\bigl(C^{L0}_{\lambda' 0,  \lambda 0}\bigr)^2
	\bigl(C^{L 0}_{j0, j'0}\bigr)^2 \notag \\
	&&\times \int_0^\infty d\eta \,  
	\gamma^{\times}_{\lambda \lambda'; j,j'}(\eta) \;
	\Big[\overline{n}_{j,j'}(\eta)+ \Theta_{j',j} \Big] \notag \\
	&&  \qquad \times \Big[\overline{n}_{j,j'}(\eta+1)+ \Theta_{j,j'} \Big] \label{eq:2_phonon_times_rate}\,,\\
	\Gamma^\prec_{j\to j'} &=&
	(2j'+1)\!
	\sum_{
		\substack{L \text{ even} \\
			\lambda \, \lambda' 
	}} \!\!
	\frac{(2\lambda+1)(2\lambda'+1)}{2L+1}
	\bigl(C^{L0}_{\lambda' 0,  \lambda 0}\bigr)^2
	\bigl(C^{L 0}_{j0, j'0}\bigr)^2 \notag \\
	&& \times \int_0^1 d\eta \,
	\gamma^{\prec}_{\lambda \lambda'; j,j'}(\eta) \;
	\Big[\overline{n}_{j,j'}(\eta)+ \Theta_{j,j'} \Big] \notag \\
	&&  \qquad \times \Big[\overline{n}_{j,j'}(1-\eta)+ \Theta_{j,j'} \Big]\label{eq:2_phonon_prec_rate} 
	\end{eqnarray}
	where $\eta$ is a dimensionless scaling parameter, which characterizes how the energy of the transition is distributed over the two phonons. $\overline{n}_{jj'}(\eta)$, and $k_{jj'}(\eta)$ are the thermal phonon number
	and the phonon momentum corresponding to the scaled transition energy $\eta E_{jj'}$.
	\begin{eqnarray}
	\gamma^{\times}_{\lambda \lambda'; j,j'}(\eta) \; &=&
	\frac{g_\text{IB}^2 |E_{jj'}|^3 }{2 \pi^3  c^4} 
	\frac{(\eta+1) \; k_{j,j'}(\eta+1)}{\sqrt{1+2(\eta+1)^2 E_{jj'}^2 \xi^2/c^2}}\notag\\
	&& 
	j_{\lambda }^2  \Bigl( r_0 k_{jj'}(\eta)\Bigr) 
	j_{\lambda'}^2  \Bigl( r_0 k_{jj'}(1+\eta)\Bigr) 
	\\
	&& \frac{\eta \; k_{j,j'}(\eta)}{\sqrt{1+2 \eta^2 E_{jj'}^2 \xi^2/c^2}}
	\big(W_{k_{j,j'}(\eta), k_{j,j'}(\eta+1)}^\times \big)^2,  \notag 
	\end{eqnarray}
	and
	\begin{eqnarray}
	\gamma^{\prec}_{\lambda \lambda'; j,j'}(\eta) \; &=&
	\frac{g_\text{IB}^2 |E_{jj'}|^3 }{4 \pi^3 c^4} 
	\frac{(1-\eta) \; k_{j,j'}(1-\eta)}{\sqrt{1+2(1-\eta)^2 E_{jj'}^2 \xi^2/c^2}}\notag\\
	&& j_\lambda^2
	\Bigl( r_0 k_{jj'}(\eta)\Bigr) j_{\lambda'}^2
	\Bigl( r_0 k_{jj'}(1-\eta)\Bigr) 
	\\
	&& \frac{\eta \; k_{j,j'}(\eta)}{\sqrt{1+2 \eta^2 E_{jj'}^2 \xi^2/c^2}}
	\big(W_{k_{j,j'}(\eta), k_{j,j'}(1-\eta)}^\prec \big)^2. \notag
	\end{eqnarray}
	Since $L$ is only summed over even numbers in eq.(\ref{eq:2_phonon_times_rate}) and (\ref{eq:2_phonon_prec_rate}), the decay is still only possible from a initial state with even (odd) $j$ to a final state with even (odd) $j'$. So the two relaxation cascades remain separated also when considering two-phonon processes.
	
	\section{Macro molecules and Landau critical rotation}
	\label{sec:Landau}
	
	As seen from Fig.\ref{fig:single-phonon} the single-phonon rotational relaxation is very different in the two cases of a usual molecule with $r_0\ll \xi$ and a macro-molecule
	$r_0>\xi$ or an atom in a shallow rotationally symmetric trap. We thus will discuss these two cases separately in the following. 
	We first consider macro molecules with a radius $r_0 > \xi $, the opposite limit is discussed in a subsequent section.

	\subsection{Relaxation rates and critical rotation}
	
	In the case of a macro-dimer the spontaneous single-phonon decay rate $\Gamma^{\textrm{1ph,sp}}_{j\to j'}$ 
	is the dominating one at low temperature and is plotted in Fig.\ref{fig:single-phonon} (b). The checkerboard pattern evolves as a consequence of the two independent relaxation cascades for even an odd angular quantum number. 
	
	As noted above, transition rates are suppressed for low angular momentum states and the molecule cannot decay to the lowest $j$ value. 
	This can be understood from analogy to linear motion of a single impurity through the condensate  \cite{Lausch2018a}. The impurity will not scatter phonons when its momentum is smaller than the Landau critical value $p_c = m_I c$ and 
	for large radii the rotation of the molecule can be approximated as a translation. 
	
	One can determine a critical angular momentum $j_c$ below which the scattering of further phonons is strongly suppressed by simultaneous energy and angular momentum
	conservation. To this end we compare the energy of two linearly moving impurities, each with momentum $p_c$,  to one rotating molecule, identifying
	\begin{equation}
	\frac{j_c(j_c+1)}{4 m_I r_0^2} = 2 \frac{p_c^2}{ 2m_I^2}.
	\end{equation}
	The corresponding Landau critical angular momentum $j_c$ is then given by
	\begin{equation}
	j_c(j_c+1) = 2 \frac{m_I^2}{m_B^2} \frac{r_0^2}{\xi^2}.\label{eq:jc}
	\end{equation}
	We note that in order to have an integer $j_c\ge 1$ the size of the molecule $r_0$ has in general to be larger than the healing length $\xi$ or we need a very heavy impurity $\mI > \mB$.
	
	Furthermore we know that a linearly moving impurity with $m_I > m_B$ can only decay into a state with momentum bigger than $p_c^{(1)} = p_c \sqrt{1-m_B^2/m_I^2}$, when only single-phonon processes are considered. In analogy to the discussion above, one can derive the minimal angular momentum $j_c^{(1)}$ a rotating macro molecule can decay into:
	\begin{align}
	j_c^{(1)}(j_c^{(1)}+1) = 2  \frac{r_0^2}{\xi^2} \left(\frac{m_I^2}{m_B^2}-1\right). \label{eq:jc1}
	\end{align}
	Both $j_c$ and $j_c^{(1)}$ fit very well to the rates calculated for the Boltzmann equation, see Fig.\ref{fig:single-phonon}(b).

	\begin{figure}[htb]
		\begin{center}
			\includegraphics[width=0.46\textwidth]{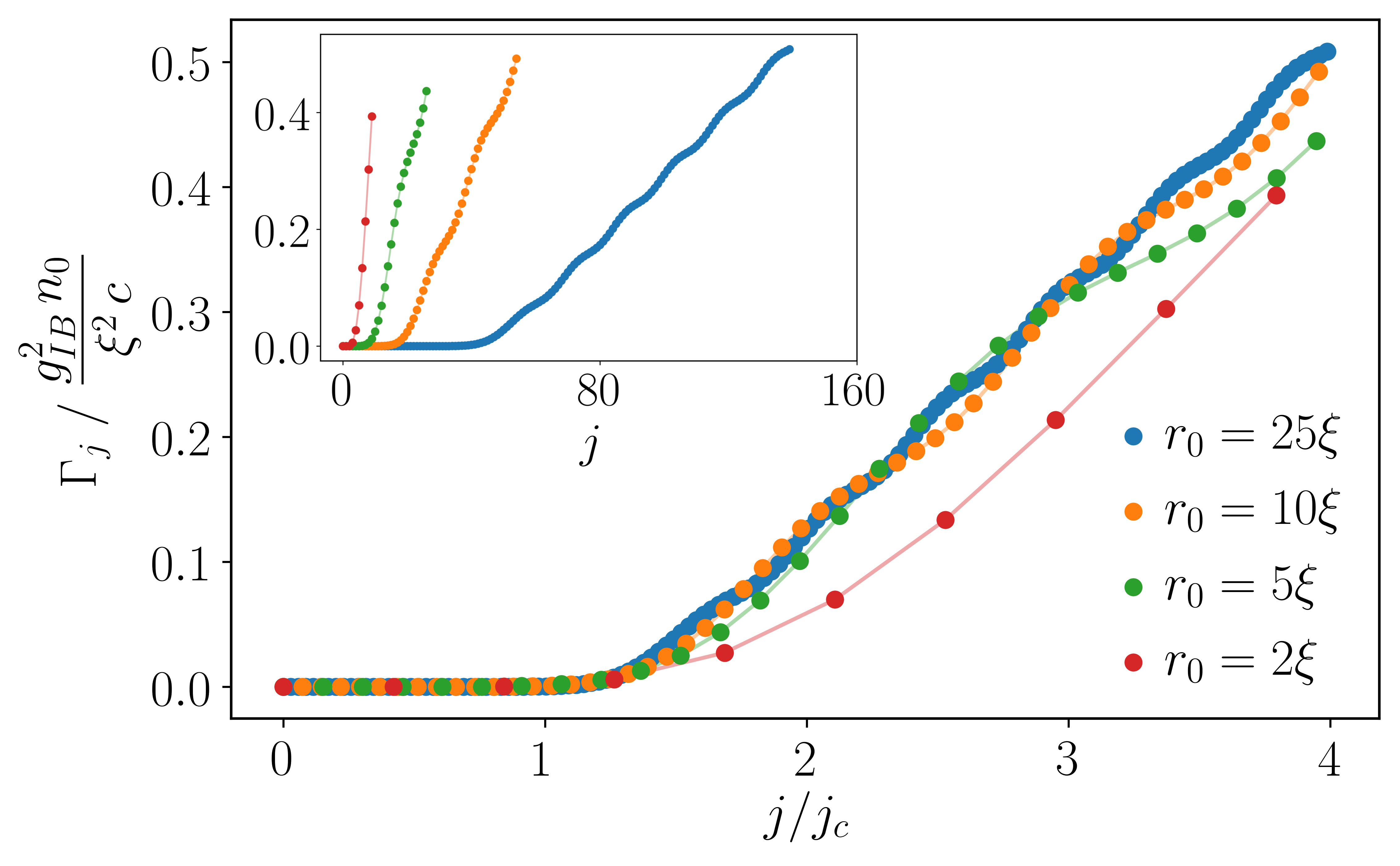}
		\end{center}
	\caption{(Color online) Single-phonon decay rate $\Gamma_{j}^\textrm{1ph,sp}$ for different molecule sizes $r_0 / \xi$, $\xi=c=1$ and $n_0\xi^3=100$.
		As shown in the inset we find non vanishing decay rates only for $j > j_c$. 
		Plotting the rates against $j$ normalized to $j_c$ the curves collapse to a single one
		if $r_0/ \xi \gg 1$.}
		\label{fig:rate-scaling}
	\end{figure}

	To verify these estimates we look at the total spontaneous single-phonon decay rate $\Gamma_{j}^\textrm{1ph,sp}$ of a molecule with angular momentum $j$, which is given by
	\begin{equation}
	\Gamma_{j}^\textrm{1ph,sp} = \sum_{j'} \Gamma_{j\to j'}^\textrm{1ph,sp}\,,
	\end{equation} 
	In the inset of Fig.\ref{fig:rate-scaling} $\Gamma_{j}^\textrm{1ph,sp}$ is plotted for different ratios $r_0/ \xi > 1$ against $j$. One clearly notices a sharp onset at $j_c$. 
	The total rates reveal oscillations that arise from projection of different spherical harmonics and more strikingly, when plotting the decay rates as function of angular momenta normalized to the critical value from eq.(\ref{eq:jc}), all curves 
	collapse to a single one when $r_0\gg \xi$. This universal  behaviour can be understood in analogy to the case of two linearly moving impurities:
	For a rotating macro molecule with angular momentum $j \gg 1$ and rotational energy equal 
	to the kinetic energy of two linearly moving impurities, each with momentum $p$, one finds
	\begin{align}
	\frac{j}{j_c} = \frac{p}{p_c},
	\end{align}
	independent on the ratio  $r_0/\xi$.

	\subsection{Cooling dynamics}

	Very similar to  \cite{Lausch2018a} one can show that the relaxation processes mediated by two-phonon processes are much slower
	than single-phonon terms in a weakly interacting 3D BEC, where $n_0\xi^3 \gg 1$, since they scale as
	\begin{equation}
	\Gamma^\textrm{2ph}/\Gamma^\textrm{1ph} \sim \bigl(n_0 \xi^3\bigr)^{-1}.
	\end{equation}
	Furthermore also thermally induced two-phonon processes 
	are very slow and not relevant below $T_c$.
	Note that the situation is markedly different in lower dimensions \cite{Lausch2018b}, where thermally-induced processes can become important
	due to the infra-red divergence of contributions by thermally occupied phonon modes.

	\begin{figure}[htb]
		\includegraphics[width=0.49\textwidth]{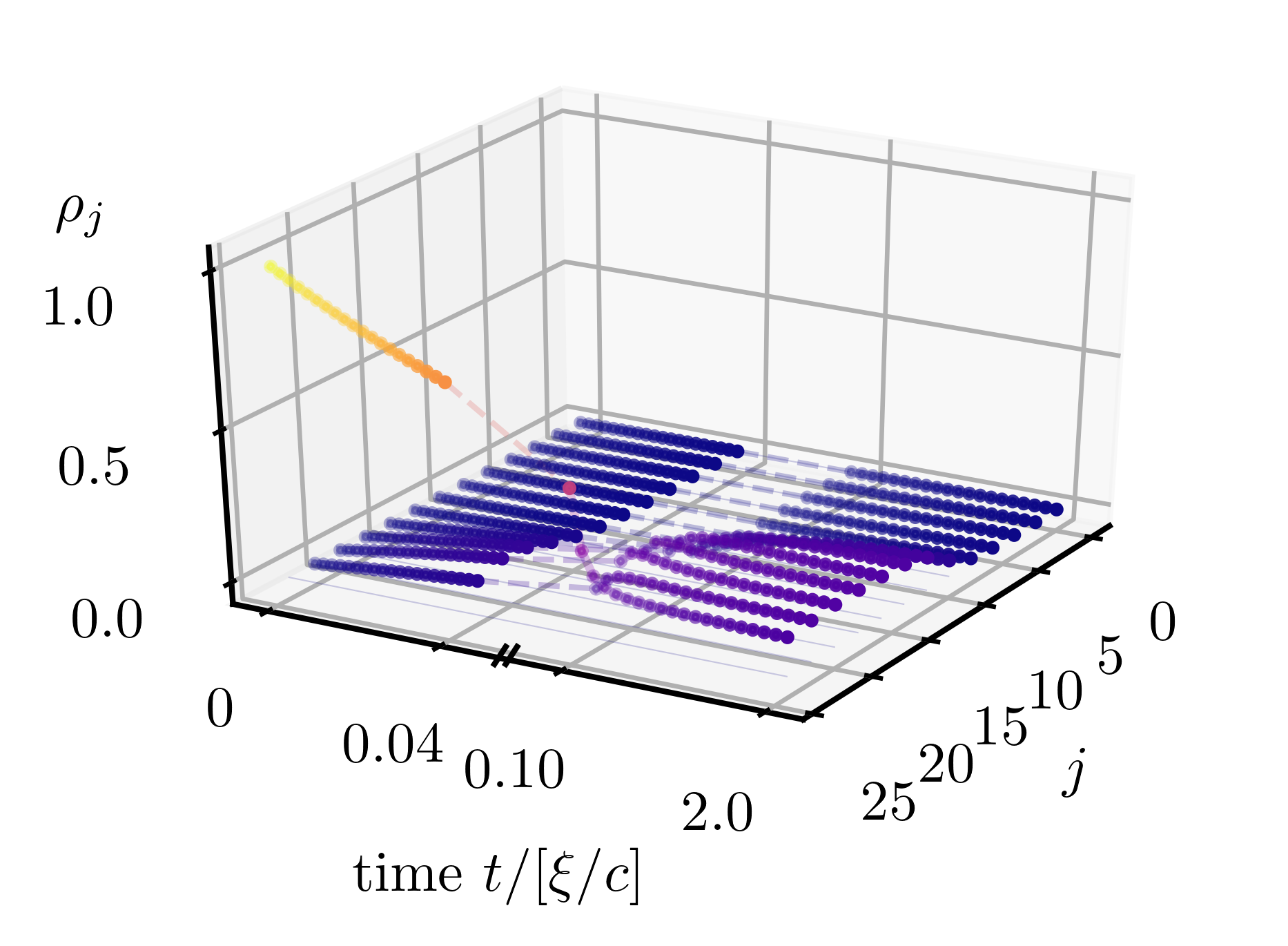}
		\caption{(Color online) Relaxation of a macro dimer initially prepared in an angular-momentum eigenstate  $\rho_{j=24} = 1$. 
			A fast approach to a pre-thermalized state with $j\ge j_c$ is clearly visible and on a longer timescale $j > j_c^{(1)}$ is populated. Here $r_0= 10 \xi$ and $m_\text{I} = 1.25 m_\text{B}$ and $j_c = 17\,(j_c^{(1)} = 10)$ and we considered a finite temperature $T = 0.01 T_c$ and density $n_0\xi^3 = 100$. }
		\label{fig:time_evolution_macro_dimer}
	\end{figure} 

	Due to the existence of a Landau critical angular momentum we expect a pre-thermalization to a non-equilibrium rotational state,
	which is visible unless $j_c \ll  j_T$, which only happens at high temperatures.  In Fig.\ref{fig:time_evolution_macro_dimer} we have plotted the
	time evolution of the occupation of angular momentum states starting at an eigenstate with $j=24$. One clearly recognizes the formation of a pre-thermalized state with $j\ge j_c^{(1)}$, while states with lower $j$ will only be populated on a much larger time scales set by two-phonon processes.
	
	We note that the mechanism of relaxation suppression discussed here is very different from that found in the opposite regime of rapidly rotating molecules in a thermal gas \cite{Stickler-PRL-2018, Milner-PRL-2014, al-Qady-PRA-2011, Forrey-PRA-2001}.

	\section{small molecules}
	\label{sec:small}

	\subsection{Single-phonon rates and angulon stability}
	
	Typical molecules have sizes much less than the healing length of the BEC $r_0 \ll \xi$.
	In this case we can drastically simplify the effective single-phonon transition rates 
	(\ref{eq:effective one phonon rates}) which yields
	\begin{equation}
	\gamma_\lambda^{jj'}=
	\begin{cases}
	\frac{c}{\pi \xi} \frac{g_{IB}^2}{g^2
	} \frac{1}{n_0 \xi^3} \frac{\xi}{r_0} \sqrt{\frac{m_B}{m_I} \Delta_{jj'}} \; j_\lambda \left(\sqrt{\frac{m_B}{2m_I} \Delta_{jj'}} \right)^2
	\quad &\lambda\enspace \text{even }\\
	0 &\text{else}
	\end{cases}  \
	\label{eq:gamma_small}
	\end{equation}
	where $\Delta_{jj'} = |j(j+1)-j'(j'+1)|$ . 
	
	Furthermore thermal contributions to the single-phonon rate can be completely disregarded as the energy spacing between adjacent rotational states
	is much larger than the thermal energy, $E_{j,j^\prime}/k_B T_c > (m_B/m_I) \bigl(n_0^{1/3} r_0\bigr)^{-2}$.
	As a consequence $ \overline{n}_{jj'}\ll 1$.

	In Fig.\ref{fig:single-phonon} (a), we plotted the transition rates $\Gamma_{j\to j'}^\textrm{1ph,sp} $ in the limit of a small molecule. An important difference to the case of a macro molecule is that the
	molecule always decays into the lowest angular momentum states $j=0$ or $1$. 
	The absence of a Landau critical rotation 
	can be understood very simply from the following argument: Phonons can resolve the rotation of the molecule if their wavelength is comparable or smaller than the molecule
	size $r_0$. Thus the relaxation is dominated by scattering of high-energy, i.e. short wavelength phonons with $k\ge r_0^{-1} \gg \xi^{-1}$. These short-wavelength phonons are
	however particle-like and there is no 
	suppression of their emission or scattering by simultaneous energy-momentum conservation. As a consequence quasi-particles arizing from the dressing of rotating molecules with angular-momentum phonons are fundamentally unstable. Furthermore in the case of a linear motion of the impurity, it is known that the transition rates are on the order of $\frac{c}{\xi} \frac{g_{IB}^2}{g^2} \frac{1}{n_0 \xi^3} $ \cite{Lausch2018a}. In contrast  eq.(\ref{eq:gamma_small}) shows that the typical transition rates for a rotating molecule are bigger by a factor $\xi/r_0$. This may raise concerns if angulons can be observed at all. However, the typical binding energies of angulons are sizable fractions of the rotational energy of the molecule. When we compare the single-phonon decay rate of angular-momentum states to the relevant energy scale, given by the rotational constant  $ B = \frac{1}{4 m_I r_0^2} $, we find
	\begin{align}
	\frac{{\Gamma}_{j\to j'}^\textrm{1ph,sp}}{B}  \propto \frac{r_0}{\xi}\ll 1
	\end{align}
	Additionally  one  recognizes from Fig. \ref{fig:rate-scaling}  that states with higher rotational number $j$ have a larger decay rate and therefore feature a broader spectral function. So while excited rotational states of a molecule in a BEC are not stable, their lifetime is still large compared to the energy of the angulon.
	
	\subsection{Thermal two-phonon contributions}
	
\begin{figure}[htb]
		\begin{center}
			\includegraphics[width=0.45\textwidth]{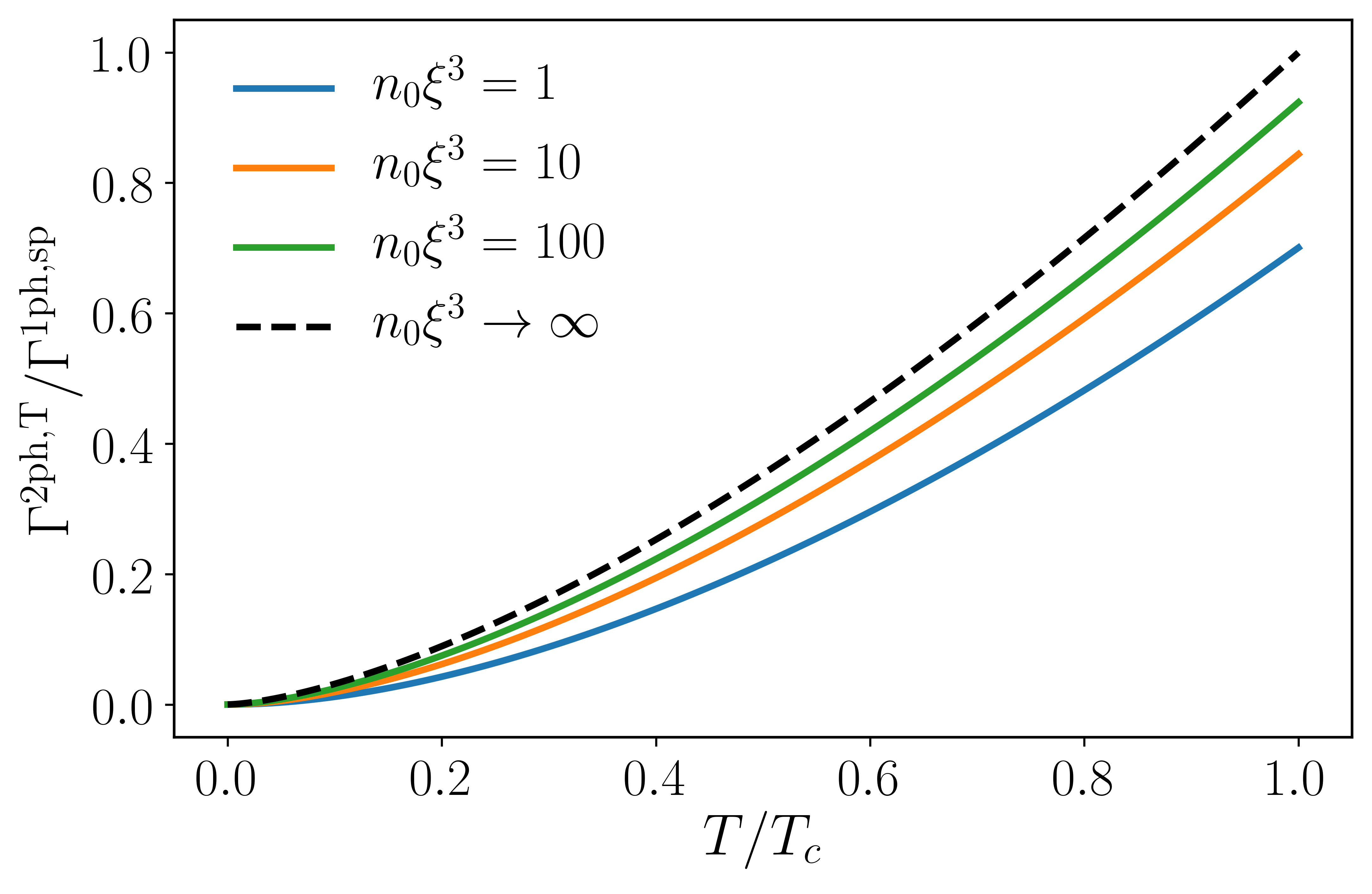}
		\end{center}
		\caption{(Color online) Ratio of thermal two-phonon to single-phonon decay rates as a function of the BEC temperature $T$ and $n_0 \xi^3 $ for small molecules.}
		\label{fig:ratio_rates_small_molecule}
	\end{figure}

	For single-phonon processes thermal effects can be neglected.  This no longer holds true for processes involving two phonons. The dominant two-phonon process is the one, were the state of the molecule decays, via absorption of a low-energy thermal phonon and subsequent (spontaneous) emission of a high energy phonon. For usually sized molecules, with $r_0 n_0^{1/3} \ll 1 $ the decay rates due to two-phonon processes are proportional to the spontaneous single phonon rates, with a proportionality factor
	which depends on the  BEC temperature  and $n_0 \xi^3$, but not on $j$ or $j'$.  
	In Fig.\ref{fig:ratio_rates_small_molecule} we have plotted the ratio of thermal two-phonon to single-phonon decay rates from numerical calculations. One recognizes that
	they approach a universal curve (dashed line) when the gas parameter $n_0 \xi^3$ increases. 
	
	As shown in the Appendix one finds
	\begin{align}
	\Gamma^{2ph, \textrm{T}}_{j \to j'} =  \Gamma^{1ph,sp}_{j \to j'} \; \frac{\sqrt{2}}{4 \pi^2} \frac{1}{n_0 \xi^3} \int_0^{\infty} d\kappa \; \xi k(\kappa)\;  \overline{n}(\kappa) .\label{eq:ratio_2_to_1_small}
	\end{align}
	Here $\kappa$ is the energy of the thermal phonon in units of $c / \xi$.  $k(\kappa)$ is the phonon momentum and $\overline{n}(\kappa)$ the thermal phonon number corresponding to this energy. For  $n_0 \xi^3\gg 1$ this expression can be further simplified which yields
	\begin{align}
	\Gamma^{2ph, \textrm{T}}_{j \to j'} \simeq  \left( \frac{T}{T_c}\right)^{3/2} \Gamma^{1ph,sp}_{j \to j'}.  
	\label{eq:T-dependence}
	\end{align}
	This simple relation holds, since the  thermal long-wavelength phonon absorbed in the two-phonon process carries effectively no angular momentum, and its energy is negligible compared to the transition energy $E_{jj'}$. 
  For a weakly interacting BEC the two processes, i.e. two-phonon scattering with absorption of a thermal phonon and single-phonon emission, only differ in that the impurity interacts with an initially condensed atom in one case and with a low-energy thermal atom in the other. Therefore the thermal contributions in the two-phonon scattering only lead to a renormalization of the single-phonon process, scaling with the thermal fraction. One recognizes, however, that at low temperature the two-phonon transition rates are still small compared to the single-phonon, so three or more-phonon processes are negligible. Furthermore direct three-body processes would not scale with the two body interaction constant $g_{ib}$ but with the three-body interaction constant, which is substantially smaller than $g_{ib}$.

	\subsection{Cooling dynamics}
	
\begin{figure} [htb]
		\includegraphics[width=0.49\textwidth]{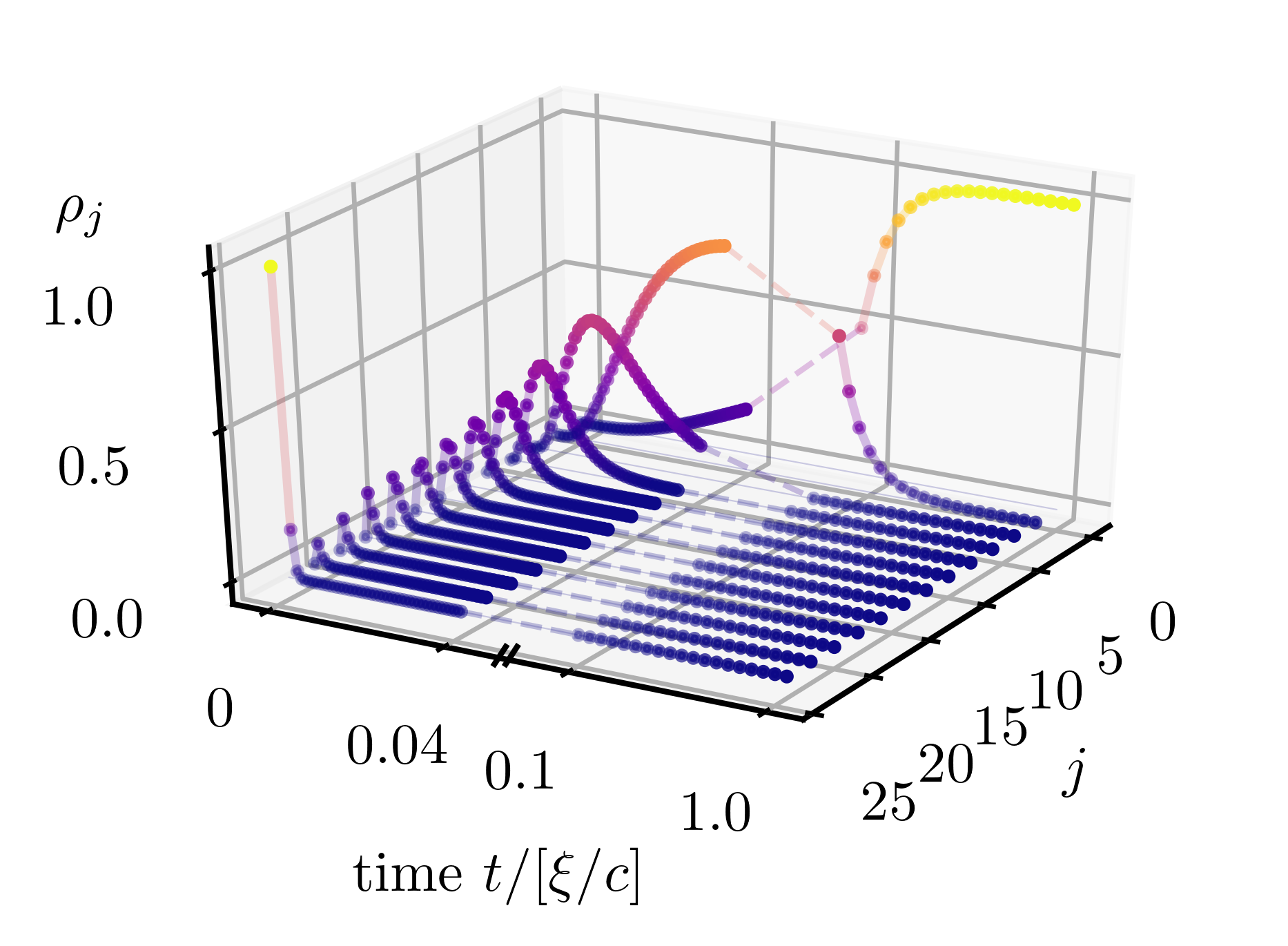}
		\caption{(Color online) Time evolution of angular momentum of a micro dimer starting at $\rho_{j=24}=1$. The color encoding matches the occupation number $\rho_j(t)$ plotted on the z-axis.  For long times $t$ we find a pumping to the final state with $j=0$. Here we have $r_0= 0.1 \xi$ and $m_\text{I} = 2 m_\text{B}$ and a finite temperature $T = 0.01 T_c$.}
		\label{fig:time_evolution_micro_dimer}
	\end{figure}

	Finally we consider also the relaxation dynamics of small molecules. To this end we solve the Boltzmann equation \eqref{eq:Boltzmann} numerically by calculating the spontaneous decay rates \eqref{eq:single-phonon-rates} and their thermal equivalent \eqref{eq:single-phonon_m_resolved_th}. In order to include two-phonon processes given in eqs.\eqref{eq:2_phonon_times_rate} and \eqref{eq:2_phonon_prec_rate} we focus on a subset of momenta up to $j\leq25$.
	Fig. \ref{fig:time_evolution_micro_dimer} shows the angular momentum decay of an initial state with $j=24$ into a final state with $j=0$. 
 For small molecules the influence of two-phonon processes increases slightly, but they do not lead to qualitative changes other than  a small modification of the  single-phonon contribution as per eq.(\ref{eq:T-dependence}). We observe a smooth and fast relaxation to a thermal state for any initial distribution of a micro dimer.

	\section{summary}
	
	We have studied the rotational relaxation of diatomic molecules immersed in a Bose-Einstein condensate of atoms
	at a temperature much below the critical value of condensation. The BEC is assumed to be weakly interacting such that a description
	in terms of a homogeneous condensate and Bogoliubov phonons is valid. The molecule was modeled as rigid rotor of two point particles.
	A more accurate description of the interaction potential between molecule and condensate atoms is possible but only affects the quantitative value of
	the coupling constants. The relaxation dynamics was analyzed with a quantum Boltzmann approach, which is valid for weak BEC impurity
	interaction. The corresponding rates can be derived from Fermi-golden rule and describe spontaneous and thermally-induced
	creation or absorption of a single phonon by the impurity out of or into the condensate as well as spontaneous and thermal two-phonon processes. The rotational
	cooling is markedly different in the case of a macro molecule with a size $r_0$ exceeding the BEC healing length $\xi$ and for a typical molecule, for which
	$r_0\ll\xi$. In the first case we found a universal behavior of the cooling rates and a Landau critical angular momentum $j_c$ caused by the
	superfluidity of the condensate in analogy to the
	case of linear motion. An initially rotationally excited molecule will quickly evolve into a pre-thermalized state which contains only angular momenta
	above a certain value $j_c^{(1)}$. The time scales of this evolution are comparable to that found in the case of linear motion. On the other hand
	for molecules of typical size, for which $r_0\ll\xi$, there is no effect of the superfluidity of the BEC since the cooling is dominated by short-wavelength
	phonons in the particle-like part of the Bogoliubov spectrum. Thus in contrast to polarons, angulons are in general not protected from decay 
	by the superfluidity of the condensate. The typical relaxation rates are much larger than in the case of macro-dimers. They are however still
	smaller than the typical binding energies of angulons.

	\section*{Acknowledgement}
	The authors would like to thank Richard Schmidt, Mikhail Lemeshko and Artur Widera for fruitful discussions. The work was supported by the German Science Foundation (DFG) within 
	SFB TR49, program number 31867626 and SFB TR185, program number 277625399.

	

	\section*{Appendix}

	In order to calculate the single-phonon transition rates eq.(\ref{eq:single-phonon_m_resolved_sp}) and (\ref{eq:single-phonon_m_resolved_th}) we first evaluate the matrix element in $\Gamma_{m\to n}= 2\pi \delta(E_m-E_n) \vert \langle m\vert H_\textrm{int} \vert n\rangle \vert^2$, which yields
	\begin{eqnarray}
	\Gamma^\textrm{1ph,sp}_{jm \to j'm'} &=& \frac{1}{2} \sum_{k\lambda\mu} \; \delta(E_{j}-E_{j'}-\omega_k) \; U_{\lambda}(k)^2 \; G^{jm^2}_{j'm',\lambda\mu}\; \\
	\Gamma^\textrm{1ph,T}_{jm \to j'm'} &= &  \frac{1}{2} \sum_{k\lambda\mu} \; U_{\lambda}(k)^2 \;\overline{n}_{k} \; \Bigl[ G^{j'm'^2}_{jm,\lambda\mu} \delta(E_{j'}-E_{j}-\omega_k)\;  \nonumber\\
	&& \qquad\quad+ G^{jm^2}_{j'm',\lambda\mu}\delta(E_{j}-E_{j'}-\omega_k) \Bigr].
	\end{eqnarray}
	We made the assumption that the phonon number $\overline{n}_{k \lambda \mu}$ depends only on $k$, which is valid for thermal phonons. The integration over the absolute value of the phonon momentum $k$ can be carried out. Furthermore by using the symmetry $G^{j'm'^2}_{jm\lambda\mu} = G^{jm^2}_{j'm'\lambda\mu}$ the thermal transition rates can be simplified. This yields
	\begin{align}
	\Gamma^\textrm{1ph,sp}_{jm \to j'm'} &= \sum_{\lambda\mu} \frac{1}{2} \frac{dk_\omega}{d\omega} U_{\lambda}(k_\omega)^2  \, \Theta_{jj'} \,  G^{jm^2}_{j'm',\lambda\mu} \Big|_{\omega = E_{jj'}} \\
	\Gamma^\textrm{1ph,th}_{jm \to j'm'} &= \sum_{\lambda\mu} \frac{1}{2} \frac{dk_\omega}{d\omega} U_{\lambda}(k_\omega)^2  \, n_{k_\omega} \,  G^{jm^2}_{j'm',\lambda\mu} \Big|_{\omega = E_{jj'}}
	\end{align}
	Where $k_\omega =\frac{1}{\xi} \sqrt{\sqrt{1+2 \omega^2 \xi^2 / c^2}-1} $ is the inverse of the dispersion relation $\omega_k$.
	The effective single phonon transition rates eq.(\ref{eq:effective one phonon rates}) are then defined as 
	\begin{equation}
	\gamma_\lambda^{jj'} = \frac{1}{2}\frac{dk_\omega}{d\omega} \,  U_{\lambda}(k_\omega)^2  \, \Big|_{\omega = E_{jj'}}
	\end{equation}
	
	The derivation of the two-phonon rates can be done in a similar way. In the following the term proportional to $U^{\times}_\lambda(k,k')$ will be considered. The derivation of the rates proportional to $U^{\prec}_\lambda(k,k')$ follows analogously. 
	When evaluating the Matrix element of $H_{int}$ proportional to  $U^{\times}_\lambda(k,k')$ one finds
	
	\begin{align}
	\Gamma^\times_{j \to j'}
	=& \frac{1}{8 \pi}\sum_{k k'\lambda \lambda'} \;  U^\times_{\lambda\lambda'}(k,k')^2  \, \bigl(\overline{n}_{k'}+1\bigr)  \notag \\
	& \quad  \overline{n}_{k} \; \delta(E_j-E_j'-\omega_k'+\omega_k)  \\ 
	& 
	\quad\sum_{\substack{LL'MM'\\\mu\mu' m'}}
	G^{LM}_{\lambda'\mu' , \lambda \mu} G^{j'm'}_{jm ,LM}
	G^{L'M'}_{\lambda'\mu' , \lambda \mu} G^{j'm'}_{jm ,L'M'} \notag 
	\end{align}
	what is already summed over all final $m'$.
	The last sums over the $D$s can be simplified using properties of the Clebsch-Gordan coefficients \cite{Varshalovich1988}.
	\begin{align}
	\sum_{\substack{LL'MM'\\\mu\mu' m'}}
	& G^{LM}_{\lambda'\mu' , \lambda \mu} G^{j'm'}_{jm ,LM}
	G^{L'M'}_{\lambda'\mu' , \lambda \mu} G^{j'm'}_{jm ,L'M'} \notag \\
	=&\; (2j'+1) \sum_L \frac{(2\lambda+1)(2\lambda'+1)}{2L+1 } C^{L0^2}_{\lambda 0 , \lambda' 0} C^{L0^2}_{j 0 , j' 0}
	\end{align}
	Furthermore the integral over $k'$ can directly be carried out and the one over $k$ is transformed into an integral over $\eta$, where $\eta E_{jj'}$ is the energy corresponding to the momentum $k$. This yields eq.(\ref{eq:2_phonon_times_rate}), where the effective transition rate is given by
	\begin{align}
	\gamma^{\times}_{\lambda \lambda'; j,j'}(\eta) \; =& \frac{|E_{jj'}|}{8 \pi } \; \frac{dk_{\omega}}{d\omega} \Big|_{\omega = \eta E_{jj'}} \,
	\frac{dk_{\omega}}{d\omega} \Big|_{\omega = (\eta+1) E_{jj'}}  \\ \notag 
	&\Bigl[U^\times_{\lambda\lambda'}\big(k_{jj'}(\eta),k_{jj'}(\eta+1) \big)\Bigr]^2
	\end{align}
	
	When a typical size molecule should be described this can be simplified further. In the following the essential steps to derive eq.(\ref{eq:ratio_2_to_1_small}), which gives the ratio between rates due to two- to single-phonon processes, are explained. Since $E_{jj'} \gg {k_b T_c} $ the molecule can not be excited, so $j>j'$. This yields
	\begin{align}
	\Gamma^{\times}_{ j \to j'} \propto \int_0^\infty d\eta \,  
	\gamma^{\times}_{\lambda \lambda'; j,j'}(\eta) \;\overline{n}_{j,j'}(\eta)
	\Big[\overline{n}_{j,j'}(\eta+1)+ 1 \Big]
	\end{align}
	The thermal phonon $\overline{n}_{j,j'}(\eta)$ number at Energy $\eta E_{jj'}$ decays exponentially fast with $\eta$. So only $\eta \ll 1$ must be considered in the integral, which is valid for $r_0 \sqrt[3]{n_0} \ll 1$.  This has the physical meaning that the energy of the thermal absorbed phonon can be neglected when compared to the energy of the emitted one. In this approximation the wavelength of the thermal phonon is much bigger than the molecule, so $ r_0 k_{jj'}(\eta) \ll1$ which leads to
	\begin{equation}
	j_{\lambda}(r_0 k_{jj'}(\eta))^2 \simeq \delta_{\lambda,0}
	\end{equation}
	Since all spherical Bessel function $j_\lambda(x)$ with $\lambda > 0$ are vanishing for small $x$. This leads to the conclusion, that the thermal absorbed phonon carries no angular momentum.
	

\end{document}